\documentclass[cameraready]{Interspeech}

\title{Context-aware child-directed speech detection from long-form recordings}

\author[affiliation={1}, orcid=0009-0004-6414-3964, equalcontribution, correspondingauthor]{Théo}{Charlot}
\author[affiliation={1}, orcid=0009-0009-2158-597X, equalcontribution]{Tarek}{Kunze}
\author[affiliation={1}, orcid=0009-0008-3193-6983]{Kaveri K.}{Sheth}
\author[affiliation={1}, orcid=0000-0003-2979-4556]{Alejandrina}{Cristia}
\author[affiliation={2}, orcid=0000-0002-6005-9368]{Marvin}{Lavechin}

\address{
    $^1$ LSCP, DEC, ENS, EHESS, CNRS, PSL University, France \\
    $^2$ Laboratoire d'Informatique et Systèmes, Université Aix-Marseille, CNRS, France
}

\email{theo.charlot@ens.psl.eu}


\usepackage{comment}
\usepackage{multirow}
\usepackage{siunitx}
\usepackage{makecell}
\usepackage{xcolor}
\usepackage{hyperref}
\hypersetup{
    colorlinks,
    linkcolor={red!50!black},
    citecolor={blue!60!black},
    urlcolor={red!50!black},
}

\newcommand{\class}[1]{\small \texttt{#1}}
\newcommand{\hp}{\hphantom{0}}

\begin{document}

\maketitle

\begin{abstract}
Automatically distinguishing child-directed speech from adult-directed speech in long-form recordings is key to scalable analyses of children's language environments. Existing approaches process utterances in isolation and have been evaluated primarily on English. We address these gaps along three dimensions. First, we fine-tune and evaluate six-self supervised models on a multilingual dataset of 182 children, showing that in-domain pre-training on child-centered recordings substantially outperforms models trained on adult speech. Second, we demonstrate that incorporating surrounding context substantially improves classification, with an absolute gain of 13.8\% in average F1-score. Third, we evaluate our model in a realistic end-to-end pipeline, from adult speech detection to addressee classification, showing that performance drops under automatic segmentation but still consistently outperforms a rule-based baseline.
\end{abstract}

\section{Introduction}

\begingroup
\renewcommand\thefootnote{}
\footnotetext{%
  \hspace{-1.8em}\fontsize{7.1}{9}\selectfont
  Code available at: \url{https://github.com/LAAC-LSCP/addressee}
}
\endgroup

Children's environments are complex, and the language input they receive is no exception. Among the sources of this complexity is the distinction between child-directed speech (CDS), the register adults typically adopt when speaking to young children, and adult-directed speech (ADS). CDS is characterized by features such as higher pitch, stretched vowels, and exaggerated intonation, as in \textit{"Look at the doggy!"}, features hypothesized to scaffold language development~\cite{saxton2009inevitability,cychosz2021acoustic,kempe2024does} Indeed, the exaggerated pitch contours of CDS may help infants segment the speech stream and identify word boundaries~\cite{stark2022word}, its acoustic saliency has been shown to preferentially capture infants' attention compared to ADS~\cite{manybabies2020quantifying}, and the use of diminutives and simplified vocabulary may help children map words onto referents more easily~\cite{yurovsky2012statistical}. Understanding the role of CDS in children's language development is therefore of considerable interest, particularly given that both the amount and nature of CDS vary substantially across cultures and socioeconomic contexts~\cite{cristia2019child,schwab2016language}. 

Child-worn microphones that record audio throughout the day have become a powerful tool for capturing naturalistic language input. However, researchers typically rely on manual annotation to distinguish between CDS and ADS, which is time and labor intensive, limiting the scale at which children's linguistic environments can be studied. Machine learning approaches could offer a scalable solution, enabling researchers to analyze large corpora that would otherwise be infeasible to annotate by hand.

The task of automatically distinguishing CDS from adult-directed speech ADS has drawn the attention of the speech processing community. It has been proposed in one of the tracks of the Interspeech 2017 Computational Paralinguistics Challenge (ComParE)~\cite{schuller17_interspeech}. The challenge organizers established a baseline using a late fusion of three systems: an end-to-end convolutional neural network with long short-term memory trained on raw waveforms, hand-crafted acoustic features, and bag-of-audio-words features, all paired with linear support vector machines, achieving a UAR of 70.2~\cite{schuller17_interspeech}. Despite several submissions~\cite{gosztolya17b_interspeech,kaya17_interspeech}, no team was able to surpass this baseline, and no open-source classifier emerged from the challenge, limiting the task's broader impact on the research community. More recently, Bang et al. (2022) trained an XGBoost classifier on LENA recordings to predict whether a 5-minute segment was predominantly CDS or ADS, achieving an AUC of 0.72~\cite{bang2022automated}. However, within such segments, CDS and ADS utterances may be interleaved, making it a coarser tool than utterance-level classification. Al Futaisi et al. (2023)  tackled the task at the utterance level by training deep neural networks on hand-crafted acoustic features using multi-task and adversarial learning strategies~\cite{al2023hearttohearta}. Their best model achieved a UAR of 67.6\% on the test set, falling short of the ComParE baseline of 70.2\%.

Despite these efforts, several limitations remain. First, most existing models have been trained predominantly on English data, representing only a narrow slice of the world's linguistic and cultural diversity~\cite{weird_2010}, a particularly consequential limitation given that the acoustic and linguistic properties of CDS are known to vary substantially across cultures and languages~\cite{cristia_cds_tismane_2019}. Second, all existing approaches train on utterances in isolation, discarding surrouding conversational context that human annotators naturally rely on when making addressee judgments. This is compounded by the fact that utterances are particularly short, with average durations of 1.5 seconds for CDS and 1.7 seconds for ADS in the ComParE training set, leaving little acoustic information for the model to exploit~\cite{schuller17_interspeech}. One may rightfully hypothesize that surrounding context (such as the presence of child vocalizations) carries useful information for addressee classification, and that models trained on isolated utterances may therefore be at a disadvantage. Third, none of the existing systems have been evaluated in a realistic setting, where speaker diarization and utterance detection must first be performed automatically on raw audio, meaning their practical utility for analyzing daylong recordings remains untested.

In this paper, we make three contributions. First, we benchmark multiple self-supervised models for CDS/ADS classification, including models classically trained on adult clean speech and two in-domain models trained on child-centered long-forms recordings, demonstrating that in-domain pretraing yields substantially better performance. Second, we show that providing the model with contextual audio windows improves classification, challenging the common practice of processing utterances in isolation and suggesting that context should be a key consideration in future model design. Third, we evaluate our approach in a fully automatic pipeline, where adult speech utterances are first detected from raw audio using the Voice Type Classifier 2.0~\cite{babyhubert_charlot2025}, before being passed to the classifier, demonstrating feasibility in a realistic deployment scenario. 

\section{Methods}

We introduce the corpora used in this study. We then formalize the addressee classification problem before introducing the self-supervised models we considered, and the context-aware fine-tuning strategy we implemented. We present the baseline against which our best model is compared and the evaluation metric. We conclude this section by providing implementation details.

\subsection{Dataset}

\noindent\textbf{\textit{Corpora.}} We gathered multiple corpora across languages and sociocultural contexts. The resulting dataset comprises a total of 22 hours of audio across 6 languages and 182 children. 

\begin{table}[ht]
 \caption{Dataset used to train and evaluate our addressee classification system. Corpora were obtained through scientific archives such as Homebank*~\cite{vandam_homebank_2016}, the Language Archive**~\cite{skilton2021ticuna}, and CHILDES***~\cite{macwhinney_childes} or accessed through direct data sharing agreements with research groups. Summary statistics are presented in Table~\ref{tab:data}.}
 \setlength{\tabcolsep}{3pt}
 \addtolength{\tabcolsep}{-0.05em}
 \fontsize{8.5pt}{8.5pt}\selectfont
 \centering
 \begin{tabular}{llSc}
     \toprule
     \textbf{Dataset} & \textbf{Language(s)} & {\makecell{\textbf{Number of} \\ \textbf{children}}} & \makecell{\textbf{Total adult} \\ \textbf{speech dur.}} \\
     \midrule
     Lyon*~\cite{lyon_homebank_2016} & French & 16 & 5h22m \\
     Cougar*~\cite{vandam2018cougar} & English & 46 & 3h32m \\
     Bergelson*~\cite{bergelson2017seedlings} & English & 20 & 2h13m \\
     Png*~\cite{png2019_cristia} & Mostly Yeli Dnye & 10 & 2h07m \\
     Lucid**~\cite{lucid_rowland_2025} & English  & 20 & 2h01m \\
     Warlaumont*~\cite{warlaumont_homebank_2024} & Mostly English  & 15 & 1h38m \\
     Soderstrom**~\cite{soderstrom_acoustical_2008} & English & 2 & 1h17m \\
     Tsimane~\cite{tsimane2017_scaff} & Tsimane  & 24 & 0h36m \\
     PhonSES~\cite{phonses_cristia_2021} & French & 6 & 0h08m \\
     Thomas***~\cite{thomas_2009} & English & 1 & 0h07m \\
     Vanuatu~\cite{vanuatu_cristia_2023} & Bislama  & 1 & 0h01m \\
     Tseltal*~\cite{tseltal_casillas_homebank_2017} & Mayan & 10 & 1h37m \\
     Winnipeg*~\cite{winnipeg_soderstrom_homebank_2016} & English  & 11 & 1h31m \\
     \midrule 
     Total & Various & 182 & 22h10m \\
     \bottomrule
 \end{tabular}
\label{tab:data}
\end{table}

\noindent\textbf{\textit{Split into partitions.}} Two corpora were held out entirely from training and reserved for evaluating cross-corpus generalization: the Tseltal corpus~\cite{tseltal_casillas_homebank_2017}, representing typologically distinct Mayan languages and a sociocultural context in which CDS is known to be infrequent~\cite{casillas2020early}, and the Winnipeg corpus~\cite{winnipeg_soderstrom_homebank_2016}, collected from English-speaking families in Canada. Together, they form our heldout set, allowing us to evaluate generalization both to an unseen language and to an unseen English corpus. 

The remaining corpora were split into training (80\%), validation (10\%), and test (10\%) splits, ensuring that no child appears in more than one split and that performance estimates are not inflated by speaker overlap.

\subsection{Addressee classification}

Addressee classification can be framed as an utterance-level classification task. Each utterance $u_i$ is a raw waveform segment associated with a ground-truth label $y_i \in\{1,\dots,K\}$ where K denotes the number of addressee classes. 

At training time, $N$ varying-length adult speech utterances (retrieved from human segmentation) are passed to the model (detailed in the next section), which outputs a predicted class probability distribution $p \in \mathbb{R}^K$. The model is trained by minimizing the categorical cross-entropy loss: 

\begin{equation}
    \mathcal{L} = -\frac{1}{N}\sum_{i=1}^{N} \log p_{y_i}
\end{equation}

\noindent where $p_{y_i} \in [0,1]$ is the predicted probability assigned to the true class $y_i$ of utterance $u_i$.

In this study, we consider $K = 3$ classes: \class{KCDS} for key child directed speech (adult speech addressed to the child wearing the microphone), \class{ADS} for adult directed speech, \class{OTHER} for speech addressed to another child, a pet, or annotated as unsure.

\subsection{Self-supervised models}

Self-supervised models pre-trained on large amounts of unlabeled speech have demonstrated strong performance across a wide range of downstream tasks, including in low-resource settings where labeled data is scarce~\cite{zhao2022improving}. This makes them particularly well-suited for addressee classification, where annotated data is limited. We experiment with six self-supervised models across three architectures: Wav2Vec 2.0 (referred to as W2V2), HuBERT, and WavLM. 

W2V2 learns speech representations via a contrastive objective over masked latent speech representations, where the model must identify the correct quantized target among a set of distractors~\cite{wav2vec2_2020}. HuBERT uses an offline clustering step to generate discrete pseudo-labels, which are then used as targets for a masked prediction objective~\cite{hubert_2021}. WavLM extends HuBERT with a denoising objective, training the model to predict masked speech representations in the presence of artificially added noise and overlapping utterances~\cite{wavlm_2022}.

We evaluate six models instantiated from these architectures. Four out-of-domain models are trained exclusively on adult speech: W2V2, HuBERT in their base configurations trained on LibriSpeech (960h)~\cite{panayotov2015librispeech}, WavLM-base-plus with the same base configuration trained on 94,000 hours of English speech, and W2V2-XLSR in its large configuration trained on 56,000 hours of multilingual speech from 53 languages~\cite{conneau21_interspeech}. The remaining two are in-domain models trained on child-centered long-forms: W2V2-LL4300 \cite{li23e_interspeech} trained on 4,300 hours of English recordings, and BabyHuBERT trained on 13,000 hours of multilingual recordings, based on W2V2 and HuBERT respectively~\cite{babyhubert_charlot2025}.

\subsection{Context-aware fine-tuning}

Addressee information may not be confined to the utterance itself: surrounding context, such as adjacent speech from other speakers or prosodic cues spanning utterance boundaries, may carry useful information for classification. This intuition is supported by a growing body of work showing that incorporating surrounding context improves performance in speech classification and recognition tasks~\cite{chang2021context,majumder2019dialoguernn,kumar2020improving}.

For an utterance of duration $d_u$, we symmetrically extend it to a total duration of $x$ seconds by adding $\frac{x - d_u}{2}$ seconds of context on each side (or give the full utterance if $d_u \geq x$). If the utterance falls at the beginning or end of the recording, context is added only on the available side. The extended audio is passed through the encoder, but only the frames corresponding to the original utterance are retained for pooling, discarding the context frames. This ensures that the context enriches the utterance representation without directly contributing to the classification decision. We experiment with multiple values for $x$ ranging from 0 (corresponding to the model receiving only the speech utterances) to 30 seconds.

\subsection{Baseline}
\label{sec:baseline}

To the best of our knowledge, no open-source supervised classifier has emerged from prior work. We therefore use as our baseline an unpublished rule-based system developed within our lab. The system operates in two steps: a voice type classifier~\cite{lavechin20_interspeech} first detects adult speech utterances, and a second step classifies adult speech as \class{KCDS} if it occurs within a fixed time threshold of the target child's vocalizations.

\subsection{Evaluation metric}

To compare the different classifiers, we use the F1-score between precision and recall, defined as:

\begin{equation*}
\text{F1-measure} = \dfrac{2 \times \text{precision} \times \text{recall}}{\text{precision} + \text{recall}}
\end{equation*}
\noindent where $\text{precision} = \text{tp} / (\text{tp} + \text{fp})$ and $\text{recall} = \text{tp} / (\text{tp} + \text{fn})$ with $\text{tp}$ the duration of true positives,  $\text{fp}$ the duration of false positives, and $\text{fn}$ the duration of false negatives.

We compute the F1-score in two flavors depending on the evaluation setting. When human-annotated boundaries are available, we compute \textit{utterance-level} F1, where tp, fp, and fn are counts of utterances. When boundaries are derived automatically, we compute \textit{frame-level} F1, where tp, fp, and fn are measured in duration, allowing us to evaluate the model in a realistic setting where segmentation errors propagate to the classifier.

\subsection{Implementation Details}

We fine-tune the self-supervised models by freezing the convolutational layers and training only the transformer layers. The final transformer layer's output is mean-pooled and projected through a linear classification head to produce logits over our three classes: \class{KCDS}, \class{ADS}, and \class{OTHER}. The predicted class is obtained via argmax. All models are trained for 10 epochs using a tri-stage learning rate schedule: a linear warmup from 1e-7 to 1e-5 over 5000 steps, a constant phase at 1e-5 for 12500 steps, and a linear decay to 5e-7 over 5700 steps. The best checkpoint is selected based on the validation performance computed at the end of each epoch. Models are trained using a batch size of 16 utterances randomly sampled from the training set. To account for training variability we train each configuration across 5 different training seeds. Training time on an A100 averages 22 min, 43 mn, and 2 h 20 min for context lengths of 0, 10, and 30 seconds, respectively.

\section{Results}

\subsection{Effect of self-supervised models}
\label{sec:exp1}

We begin by addressing our first question, comparing multiple self-supervised models fine-tuned on our addressee classification task (Table \ref{tab:exp1}).

\begin{table}[htbp]
    \fontsize{8pt}{8pt}\selectfont
    \centering
    \caption{Validation F1-scores (\%) and respective standard deviations (\%) over 5 runs  for different self-supervised models ($\uparrow$ higher is better). \class{KCDS}, \class{ADS}, and \class{OTHER} refer to key-child-directed, adult-directed, and other speech. Ave. is the macro-average F1-score across speakers. All models use their base architecture except W2V2-XLSR (large). Models marked with $^\dagger$ were pretrained on child-centered recordings 
(child and adult speech); all others were pretrained on adult speech only. Best results are reported in bold.}
    \label{tab:exp1}
    \addtolength{\tabcolsep}{-0.2em}
    \begin{tabular}{@{}lcccc@{}}
        \toprule
         & \multicolumn{4}{c}{F1-scores (\%) $\uparrow$} \\
        Model & \class{KCDS} & \class{ADS} & \class{OTHER} & Ave. \\
        \midrule
        W2V2             & 72.5 $\pm$ 0.5 & 49.2 $\pm$ 2.7 & 14.8 $\pm$ 2.2 & 45.5 $\pm$ 1.3 \\
        HuBERT           & 73.1 $\pm$ 0.3 & 54.2 $\pm$ 1.8 & 14.7 $\pm$ 3.1 & 47.3 $\pm$ 1.1 \\
        WavLM            & 70.7 $\pm$ 1.0 & 51.7 $\pm$ 1.8 & \hp2.0 $\pm$ 4.4 & 41.5 $\pm$ 1.7 \\
        W2V2 XLSR        & 72.1 $\pm$ 0.3 & 53.0 $\pm$ 1.2 & 14.8 $\pm$ 1.4 & 46.6 $\pm$ 0.5 \\
        W2V2 LL4300 $^\dagger$      & 74.4 $\pm$ 0.3 & 54.4 $\pm$ 2.1 & \hp9.4 $\pm$ 3.9 & 46.1 $\pm$ 1.3 \\
        BabyHuBERT $^\dagger$       & \textbf{77.4} $\pm$ 0.2 & \textbf{62.0} $\pm$ 0.9 & \textbf{20.1} $\pm$ 2.7 & \textbf{53.2} $\pm$ 1.1 \\
        \bottomrule
    \end{tabular}
\end{table}

Among out-of-domain models pre-trained on adult speech, W2V2, HuBERT and W2V2-XLSR achieve comparable F1-scores in the 45\% - 57\% range, suggesting no clear benefit from multilingual pretraining on a substantially larger corpus (56,000h for W2V2 XLSR vs. 960h for W2V2 and HuBERT). WavLM, despite its denoising pretraining objective, stands out as an outlier with a near-zero F1-score on the Other class.

Among in-domain models, W2V2-LL4300 surprisingly fails to outperform out-of-domain models despite being trained on 4,300 hours of child-centered long-forms. BabyHuBERT, on the other hand, achieves the best overall performance with an average F1-score of 53.2$\pm$1.1\%, which may be attributed to its pretraining on 13000 hours of multilingual child-centered long-forms. Across all models, the Other class consistently yields the lowest F1-scores, likely due to its heterogeneous composition. Indeed, Other encompasses subcategories such as pet-directed and other-child-directed speech that are acoustically similar to \class{KCDS}, making discrimination particularly challenging.

Overall, while acknowledging the many counfouding factors across models (architecture, pre-training objective, implementation, etc.), our results suggest that in-domain pre-training may be an important driver of performance for addressee classification. Models pretrained on child-centered recordings are indeed exposed to the very acoustic conditions that characterize the downstream task, which differ substantially from the clean adult speech found in standard pretraining corpora.

\subsection{Effect of context duration}
\label{sec:exp2}

We now investigate whether incorporating surrounding context improves addressee classification, and if so, what context duration is optimal. 

\begin{figure}[htbp]
    \centering
    \includegraphics[width=.9\linewidth]{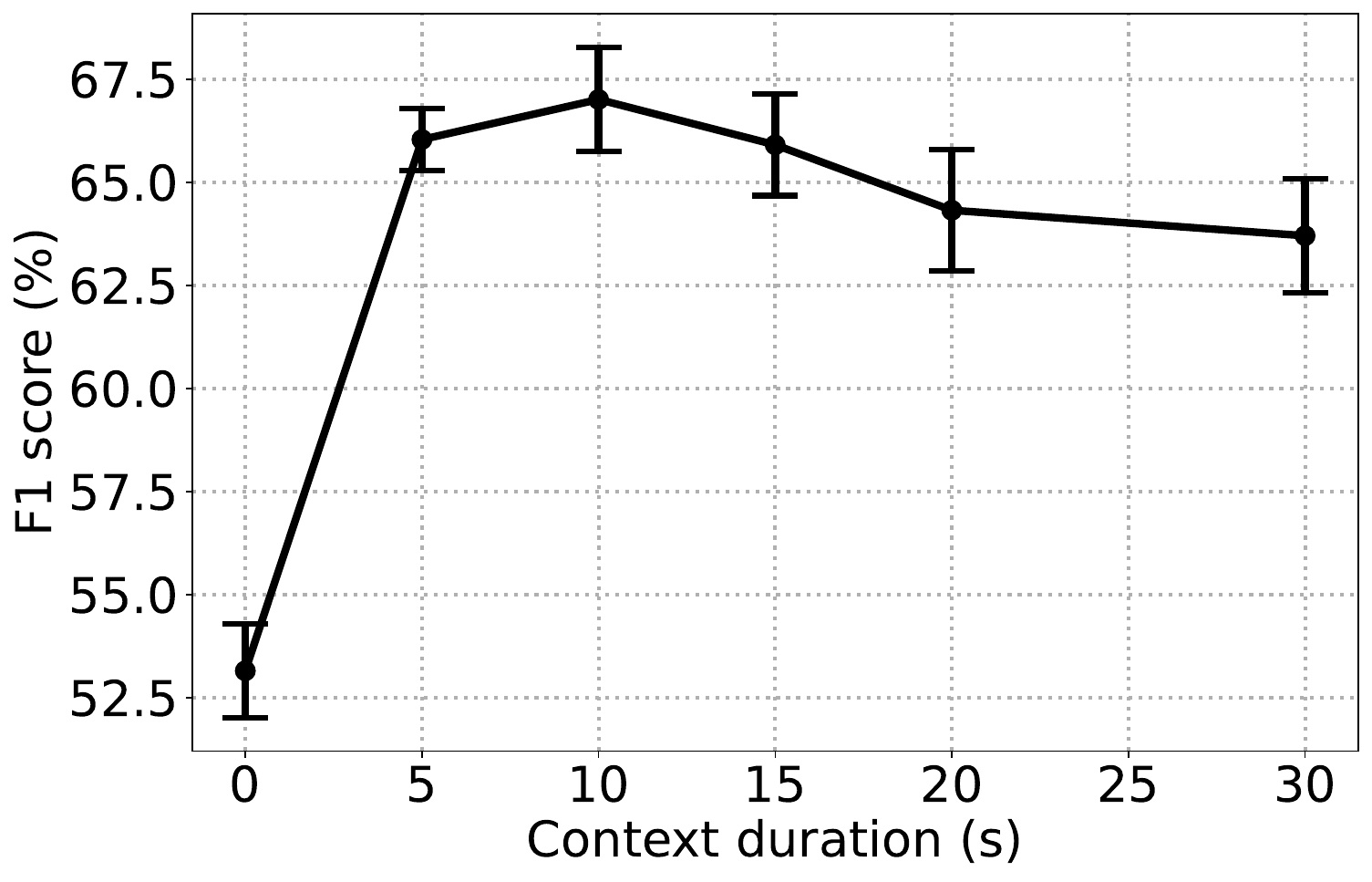}
    \caption{Effect of context duration on the validation F1-score (\%), where 0 seconds corresponds to no context, i.e., the model receives only the target utterance. Standard deviations are computed across 5 random seeds.} 
    \label{fig:exp2}
\end{figure}

As shown in Figure \ref{fig:exp2}, adding context yields a substantial improvement over the more standard no-context condition, with F1-score jumping from 53.2$\pm$1.1\% at 0 seconds to 67.0$\pm$1.3\% at 10 seconds. Performance then gradually declines beyond 10 seconds until it reaches 63.7$\pm$1.4\% at 30 seconds. Although it is hard to provide precise evidence as to why, we hypothesize that longer contexts dilute the relevant signal, causing the model to rely on cues that do not generalize beyond training set. 

Overall, our results confirm that context is highly beneficial for addressee classification. Surrounding speech carries informative cues that are absent from the target utterance alone, such as knowing who spoke just before or after it. Simply considering contextual windows of 10 seconds yields a massive absolute improvement of 13.8\% in terms of average F1-score over the no-context condition. On the test set, the best-performing model (trained with 10-second contextual windows) achieves F1-scores of 81.6\%, 78.9\%, and 36.2\% for \class{KCDS}, \class{ADS}, and \class{Other} respectively, confirming that validation performance is stable and not an artifact of hyperparameter selection. We refer to this model as BabyHuBERT-addressee in the remaining analysis.

\subsection{Assessing performance in realistic conditions}
\label{sec:exp3}

So far, our analyses have relied on human-annotated speech boundaries, which does not reflect how a fully automatic pipeline would perform in realistic conditions. We now turn to the heldout set to assess performance under more realistic conditions, where adult speech utterances are obtained automatically using VTC 2.0~\cite{babyhubert_charlot2025}. Given that utterances do not match up across human vs automated diarization, we switch here to a window-level F1-scores with windows of $100\text{ms}$ where the true positives, false positives, and false negatives are computed in terms of classified windows. Since the rule-based baseline does not distinguish \class{ADS} from \class{OTHER}, we merge these two classes into a single overheard speech class (\class{OHS}) to allow for comparison.

\begin{table}[ht]
    \fontsize{9pt}{9pt}\selectfont
    \centering
    \caption{Frame-level F1-scores (\%) on the heldout set for the rule-based baseline and BabyHuBERT-addressee using either human or automated VTC 2.0 segmentation ($\uparrow$ higher is better). \class{KCDS} and \class{OHS} refer to the key-child-directed and overheard speech. Ave. is the macro average F1-score across speakers. Best results are reported in bold.}
    \label{tab:exp3}
    \begin{tabular}{@{}lccc@{}}
        \toprule
         & \multicolumn{3}{c}{F1-scores (\%) $\uparrow$}  \\
        Model & \class{KCDS} & \class{OHS}  & Ave. \\
        \midrule
        \multicolumn{4}{c}{\textit{Human segmentation}} \\
        Rule-based           & 37.9          & 32.3          & 35.1 \\
        BabyHuBERT-addressee & \textbf{64.1} & \textbf{84.1} & \textbf{74.1} \\
        \midrule
        \multicolumn{4}{c}{\textit{VTC 2.0 segmentation}} \\
        Rule-based           & 36.3          & 14.9          & 25.6 \\
        BabyHuBERT-addressee & \textbf{43.4} & \textbf{33.8} & \textbf{38.6} \\
        \bottomrule
    \end{tabular}
\end{table}

Table~\ref{tab:exp3} compares BabyHuBERT-addressee against the rule-based baseline (Section \ref{sec:baseline}) using either human or automatically generated speech boundaries. As expected, the rule-based baseline performs poorly, as temporal proximity to the target child is a weak proxy for addressee classification compared to direct acoustic modeling. BabyHuBERT-addressee consistently outperforms the baseline in both conditions, with absolute improvements of 39\% and 13\% in average F1-score for human and VTC 2.0 segmentation respectively. There is a substantial difference in performance between the two held-out datasets (not shown in Table~\ref{tab:exp3} ), with average F1-scores of 80.8\% for Winnipeg and 58.3\% for Tseltal. The lower performance on Tseltal is likely due to its challenging recording conditions: noisy outdoor rural settings and a higher proportion of ADS from the frequent presence of non-parent speakers. Importantly, performance drops substantially (by 35.5\% in average F1-score) when switching from human to automatic segmentation, suggesting that segmentation errors introduced by VTC 2.0 propagate through the pipeline and compound with addressee classification errors.

\section{Discussion and conclusion}

Our results show that large-scale, automatic detection of \textit{who speaks to the child} from naturalistic long-form recordings is feasible. Importantly, our results highlight two key factors for improving performance: domain-matched multilingual pretraining, with BabyHuBERT consistently outperforming other self-supervised models, and the incorporation of context, with a 10-second contextual window yielding an absolute improvement of 13.8\% over the no-context condition. 

A key limitation of prior work has been the absence of open-source tools for addressee classification, which has hindered cumulative progress on the task. By releasing our model and code, we hope to provide the community with a shared baseline that can be be built upon, enabling more systematic comparisons across corpora. We believe that reproducibility and open-sourcing are critical for advancing the field, particularly given the diversity of recording conditions and sociocultural contexts that a robust classifier must handle.

On the modeling side, our context-aware fine-tuning encodes the full contextual window through the transformer, which comes at a significant computational cost, with training time increasing from 22 minutes to over 2 hours when moving from 0 to 30 seconds of context. More efficient alternatives could include hierarchical architectures where the target utterance and its context are encoded separately, and a cross-attention module then allows the target utterance to selectively retrieve relevant information from the context frames, avoiding the need to pass the full concatenated audio through the transformer, as explored in other speech tasks~\cite{chang2021context,majumder2019dialoguernn}. Such architectures would also lend themselves to exploring asymmetric context windows, where preceding and following speech contribute differently to the classification decision.

On the developmental science side, our work is a first step towards cross-liguistic quantification of children's input, making it possible to examine how the amount and distribution of child-directed speech may vary across communities and developmental contexts without relying exclusively on manual annotation. A natural important next step would be to assess whether automatically-derived estimates of child-directed and overheard speech correlate with human-annotated estimates at the recording level, and whether such correlations hold equally across diverse sociocultural contexts. Such validation would be critical before deploying our system at scale for language development research purposes. 

\clearpage

\section{Acknowledgments}

This work was performed using HPC resources from GENCI-IDRIS (Grant 2024-AD01101545 and 2025-AD011016414) and was supported in part by the Agence Nationale pour la Recherche (ANR-17-EURE-0017 Frontcog, ANR10-IDEX-0001-02 PSL). TC was funded by an ERC grant (InfantSimulator, 101142705); AC, KS and TK were funded by an ERC grant (ExELang, 101001095). ML acknowledges funding from Simons Foundation International (034070-00033). Views and opinions expressed are those of the authors only and do not necessarily reflect those of the European Union or the European Research Council. Neither the European Union nor the granting authority can be held responsible for them.

\begingroup
    \hypersetup{hidelinks}
    \bibliographystyle{IEEEtran}
    \bibliography{refs}
\endgroup
\end{document}